\documentclass[final, a4paper journal, twocolumn]{IEEEtran}

\usepackage{threeparttable} 
\usepackage{graphicx}
\usepackage{picinpar}
\usepackage{rotating}

\usepackage{psfrag}
\usepackage{placeins}
\usepackage{times}
\usepackage{amsmath}
\usepackage{amsfonts}
\usepackage{subfigure}
\usepackage{cite}
\usepackage{color}

\renewcommand{\baselinestretch}{1.0}

\usepackage{pifont}
\usepackage{amssymb}
\usepackage{setspace}
\usepackage{changebar}
\usepackage{flafter}
\DeclareGraphicsExtensions{.eps,.pdf}
\usepackage{booktabs}
\usepackage{tabularx}
\usepackage{array}

\newcommand{\PreserveBackslash}[1]{\let\temp=\\#1\let\\=\temp}
\newcolumntype{C}[1]{>{\PreserveBackslash\centering}p{#1}}
\newcolumntype{R}[1]{>{\PreserveBackslash\raggedleft}p{#1}}
\newcolumntype{L}[1]{>{\PreserveBackslash\raggedright}p{#1}}

\usepackage{mdwmath}
\usepackage{mdwtab}
\usepackage{dblfloatfix}
\usepackage{morefloats}
\usepackage{url}

\begin{document}
\bibliographystyle{../../../shaoshi_bib/IEEEtran}
\title{Large-Scale MIMO is Capable of Eliminating Power-Thirsty Channel Coding for Wireless Transmission of HEVC/H.265 Video}
\author{Shaoshi~Yang,~\IEEEmembership{Member,~IEEE}, Cheng~Zhou, Tiejun~Lv,~\IEEEmembership{Senior~ Member,~IEEE},~Lajos~Hanzo,~\IEEEmembership{Fellow,~IEEE}
\thanks{

The financial support of the European Research Council's Advanced Fellow Grant and of National Natural Science Foundation of China (NSFC) (Grant 61201268, 61471400) is gratefully acknowledged.  

S. Yang is with the School of Electronics and Computer Science, University of Southampton, Southampton SO17 1BJ, UK (e-mail: sy7g09@ecs.soton.ac.uk).

C. Zhou is with the Hubei Key Laboratory of Intelligent Wireless Communications, School of Electronics and Information Engineering, South-Central University for Nationalities, Wuhan 430074, China. He was also with the School of Electronics and Computer Science, University of Southampton,
Southampton SO17 1BJ, UK (e-mail: czhou@mail.scuec.edu.cn).

T. Lv is with the School of Information and Communication
Engineering, Beijing University of Posts and Telecommunications, Beijing 100876, 
China (e-mail: lvtiejun@bupt.edu.cn).%

L. Hanzo is with the School of Electronics and Computer Science, University of Southampton,
Southampton SO17 1BJ, UK (e-mail: lh@ecs.soton.ac.uk).
}}

\markboth{published on IEEE Wireless Communications, vol. 23, no. 3, pp. 57-63, June 2016}%
{Shell \MakeLowercase{\textit{et al.}}: Bare Demo of IEEEtran.cls
for Journals}

\maketitle
\begin{abstract}
A wireless video transmission architecture relying on the emerging large-scale multiple-input--multiple-output (LS-MIMO) technique is proposed. Upon using the most advanced High Efficiency Video Coding (HEVC) (also known as H.265), we demonstrate that the proposed architecture invoking the low-complexity linear zero-forcing (ZF) detector and dispensing with any channel coding is capable of significantly outperforming the conventional small-scale MIMO based architecture, even if the latter employs the high-complexity optimal maximum-likelihood (ML) detector and a rate-$1/3$ recursive systematic convolutional (RSC) channel codec. Specifically, compared to the conventional small-scale MIMO system, the effective system throughput of the proposed LS-MIMO based scheme is increased by a factor of up to three and the quality of reconstructed video quantified in terms of the peak signal-to-noise ratio (PSNR) is improved by about $22.5\, \text{dB}$ at a channel-SNR of $E_b/N_0 \approx 6\,\text{dB}$ for delay-tolerant video-file delivery applications, and about $20\,\text{dB}$ for lip-synchronized real-time interactive video 
applications. Alternatively, viewing the attainable improvement from a power-saving perspective, a channel-SNR gain as high as $\Delta_{E_b/N_0}\approx 5\,\text{dB}$ is observed at a PSNR of $36\, \text{dB}$ for the scenario of delay-tolerant video applications and again, an even higher gain is achieved in the real-time video application scenario. Therefore, we envisage that LS-MIMO aided wireless multimedia communications is capable of dispensing with power-thirsty channel codec altogether!
\end{abstract}

\begin{IEEEkeywords}
Large-scale/massive MIMO, HEVC/H.265, H.264, wireless video transmission, channel coding 
\end{IEEEkeywords}

\makeatletter
\def\hlinewd#1{%
  \noalign{\ifnum0=`}\fi\hrule \@height #1 \futurelet
   \reserved@a\@xhline}
\makeatother

\IEEEpeerreviewmaketitle

\section{Introduction}
\IEEEPARstart{T}{he} demand for video traffic in wireless networks is set to soar in the near future\cite{Cisco_2014:whitepaper_mobile_data_growth_forcast}. To support improved wireless video services,  two major technological directions may be pursued, namely improving the video source compression efficiency and increasing the wireless channel capacity. 

In the first direction, it has been shown\cite{Sullivan_2012:HEVC_overview, Pourazad_2012:HEVC_magazine} that the emerging High Efficiency Video Coding (HEVC/H.265) standard\cite{HEVC_2013:ITU} is capable of significantly improving the video compression efficiency at a given reconstructed video quality compared to the existing H.264 standard\cite{Wiegand_2003:H_264_standard, Hanzo_2007:video_book}, albeit at the cost of an increased computational complexity\cite{Sullivan_2012:HEVC_overview, Pourazad_2012:HEVC_magazine}. However, an increased compression efficiency usually makes the coded video stream more vulnerable to packet losses \cite{Pourazad_2012:HEVC_magazine, Yongkai_2015:video}. The traditional approach of compensating  this side effect is to employ sufficiently powerful error correction codes (ECC), such as recursive systematic convolutional (RSC) codes, turbo codes or low-density parity-check (LDPC) codes for 
improving the bit/packet error rate (BER/PER)\cite{Yongkai_2015:video, Vetro_2005:video_error_wireless}. Unfortunately, the redundant bits introduced by ECC not only require additional energy to transmit, which is a serious burden for low-power devices such as mobile terminals and wireless sensors, but also decrease the effective transmission rate of a system. For example, when a rate-1/3 ECC is used, the number of bits to be transmitted is three times as high as that of the system dispensing with ECC. \textit{Therefore, how to strike an attractive tradeoff amongst the transmission reliability, the video transmission rate and the energy efficiency is an important research problem.}

In the second direction, as one of the key technologies for future 5G wireless communications, large-scale multiple-input multiple-output (LS-MIMO) techniques have been shown to be capable of providing substantial transmission rate, reliability and/or energy efficiency improvements through the use of a large number of base station (BS) antennas \cite{Marzetta2010:massive_MIMO, Rusek_2013:massive_MIMO, Anzhong_2014:massive_MIMO_source_localization, Shaoshi_2015:MIMO_detection_survey}. 

Against the above background, in this paper we present a new system architecture based on the interdisciplinary study of how to transmit HEVC/H.265 compressed video over wireless channels with the aid of LS-MIMO techniques. Our novel contributions are summarized as follows. 
\begin{itemize} 
\item 
 we devise a low-complexity architecture for multipoint-to-point (M2P) H.265 video transmission relying on an uplink LS-MIMO system. Upon using a moderate number of BS antennas and a low-complexity linear zero-forcing (ZF) LS-MIMO detector, we demonstrate that the proposed architecture dispensing with any ECC is capable of offering an improved performance in terms of both the achievable transmission rate and the reconstructed video quality, compared with its traditional small-scale MIMO counterpart that relies both on a sophisticated ECC codec and on the optimal high-complexity maximum-likelihood (ML) symbol detector.  
\item To the best of our knowledge, using LS-MIMOs for wirelessly transmitting the most advanced HEVC/H.265 video stream that is traffic-demanding has never been reported before in the open literature. Our results explicitly demonstrate that at the expense of an increased number of antenna elements and radio frequency (RF) chains, our much simpler signal processing architecture is capable of significantly outperforming the higher-complexity, higher-delay, higher-energy-consumption traditional architecture. The significance of our results is that they potentially open up a new \textit{interdisciplinary} research direction with the objective of identifying hitherto unexplored reduced-complexity system architectures, which are facilitated by the cutting-edge LS-MIMO combined with the emerging HEVC/H.265 video processing. 
\end{itemize}
\section{A Case-Study of the Benefits of LS-MIMO Systems}\label{sec: case_study}
\begin{figure}[tbp]
\centering{\includegraphics[width=3.5in]{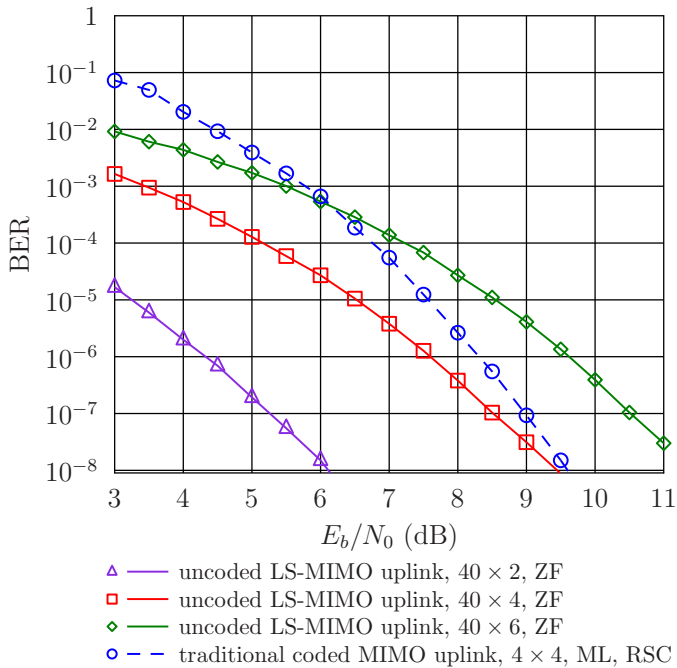}}
\caption{BER versus the channel-SNR $E_b/N_0$ for an uncorrelated flat Rayleigh fading channel.} \label{bervssnr}
\end{figure}
In order to demonstrate the benefits of LS-MIMOs, let us commence with a pair of parallel case studies concerning the BER performance of an uplink LS-MIMO system and of a traditional small-scale uplink MIMO system. The energy efficiency of both systems is also discussed.

Both systems are assumed to be synchronous and may be characterized by the classic MIMO system model of ${\bf y} = {\bf Hs} + {\bf n}$, 
where ${\bf y} \in \mathbb{C}^{N_r \times 1}$ is the signal vector received at the BS in each time instance; ${\bf H} \in \mathbb{C}^{N_r\times N_t}$ denotes the random wireless channel matrix between $N_t$ single-antenna users and the BS equipped with $N_r$ antennas; and ${\bf s} \in \mathbb{C}^{N_t \times 1} $ represents the symbol vector transmitted in each time instance, whose elements $s_i$, $i = 1, 2, \cdots, N_t$, are randomly taken from a modulation constellation $\mathbb{A}$ and satisfy the total transmit power constraint of $\text {E} (\lVert {\bf s}\rVert^2) = 1$. In other words, the total transmit power of all users at each time instance is assumed to be constant and the transmit power of each single-antenna user in our LS-MIMO system is proportional to $1/N_t$. Additionally, ${\bf n} \in \mathbb{C}^{N_r \times 1}$ denotes the complex-valued additive white Gaussian noise vector (AWGN), whose elements obey the independent identically distributed (i.i.d) $\mathcal {CN}(0, 2\sigma^2)$. We assume an uncorrelated flat Rayleigh fading channel, which implies that each entry of $\bf H$, i.e. $h_{ji}$, is a complex-valued Gaussian variable. In order to ensure that the site-specific received-SNR biases caused by large-scale fading such as pathloss and shadowing are removed, the channel coefficients $h_{ji}$ 
are normalized as  $\mathcal {CN} (0,1)$.  
     
\subsection{The Advocated LS-MIMO Configuration} For the uplink LS-MIMO, $N_r \gg N_t $ is required. Hence, in the uplink LS-MIMO system considered, $N_t$ was set to $N_t= \{2, 4, 6\}$ and $N_r$ was set to the moderate value of 40, whilst in most LS-MIMO literature\cite{Marzetta2010:massive_MIMO, Rusek_2013:massive_MIMO, Anzhong_2014:massive_MIMO_source_localization,  Shaoshi_2015:MIMO_detection_survey} hundreds of BS antennas are typically assumed. As a result, the uplink transmit power of each user in our LS-MIMO system becomes 1/2, 1/4 and 1/6 of the total transmit power, respectively. The low-complexity linear ZF based MIMO detector of\cite{Shaoshi_2015:MIMO_detection_survey} was invoked at the BS, but no ECC was employed. The linear ZF detector is characterized by
$\hat {\bf s} = \text {det}\left( {\bf Wy}\right)$, 
where $\text {det} (\cdot)$ denotes the hard-decision detector implemented by a slicer, and we have ${\bf W} = ({\bf H}^H{\bf H})^{-1}{\bf H}^H $ for the scenario of $N_r > N_t $, while ${\bf W} = {\bf H}^{-1}$ if $N_r =  N_t$. 
  
Notably, for the uplink LS-MIMO systems it has been shown in \cite{Larsson_2013:EE_SE_massive_MIMO} that when the number of BS antennas $N_{r} $ is sufficiently large, the transmit power of each single-antenna user can be made proportional to $1/N_{r}$, provided that the BS has the perfect knowledge of the channel state information (CSI) $\bf H$; alternatively, it becomes  proportional to $1/\sqrt{N_{r}}$, if $\bf H$ is estimated from the uplink pilots, with no performance degradation in both cases. This is indeed a more optimistic result than the $1/{N_t}$-power-scaling assumption we made above. Anyway, in LS-MIMO systems each antenna element uses an extremely low transmit power, which could be on the order of milliwatts. Nonetheless, in practice, several realistic imperfections  will undoubtedly prevent us from fully realizing the above optimistic power savings, owing to the associated CSI estimation errors, the interference imposed by other cells and the additional RF front end circuitry. The undersized impact of these factors on the system's energy consumption is highly dependent on the specific system implementations, which is hence yet to be rigorously investigated. Even so, the prospect of 
saving an order of magnitude in transmit power is significant, 
because we can achieve an improved system performance under the 
same regulatory power constraints, and because 
the escalating energy consumption of both the BSs and the users is a growing 
concern  \cite{Larsson_2013:EE_SE_massive_MIMO, Rusek_2013:massive_MIMO, Kent_2013:EE_OFDMA_SISO_single_cell}.

\subsection{The Benchmark Small-Scale MIMO Configuration} Additionally, a RSC-aided conventional $(4\times4)$-element small-scale spatial-multiplexing MIMO system was used as the benchmark. Therefore, the uplink transmit power of each user becomes 1/4, which indicates that both the per-user transmit power and the total transmit power of the two MIMO systems are comparable. The generator polynomials of $[133,165,171]_{8}$ represented in octal form were employed for configuring the rate-$1/3$ RSC code, which was decoded at the receiver by the classic soft-output Viterbi algorithm. Additionally, the optimal ML based MIMO detector of\cite{Shaoshi_2015:MIMO_detection_survey} was used, which relies on a high-complexity ``brute-force search'' formulated as ${\hat {\bf{s}}} = \arg \mathop {\min }\limits_{{\bf{s}}
\in \mathbb{A}^{N_t}} {\rm{ }}  \left\| {{\bf{y}} - {\bf{Hs}}}
\right\|^2$. 
Note that there is no iteration between the MIMO detector and the channel decoder. 
We assume that quadrature amplitude modulation (QAM) signals were transmitted in both systems. In Fig. \ref{bervssnr} we present the BER versus $E_b/N_0$ performance of both systems invoking 4-QAM. Observe from Fig. \ref{bervssnr} that the BER performance of our uncoded uplink LS-MIMO system becomes even better than that of the RSC-aided traditional small-scale MIMO system, when $N_t$ of the former scheme is no more than 4. 

Compared to that of the small-scale MIMO benchmark system,  the energy consumption of the signal processing at the LS-MIMO scheme's transceiver is expected to be significantly reduced, since the high-complexity channel codec is eliminated and the MIMO detector is the low-complexity ZF detector. By contrast, the small-scale MIMO benchmarker considered has to use a sophisticated channel codec and the optimal ML MIMO detector, which consumes more power and imposes a higher delay, and yet only provides a less attractive overall performance. 

\section{The Proposed Video Transmission System Architecture}
\begin{figure}[t]
\centering{\includegraphics[width=3.5in]{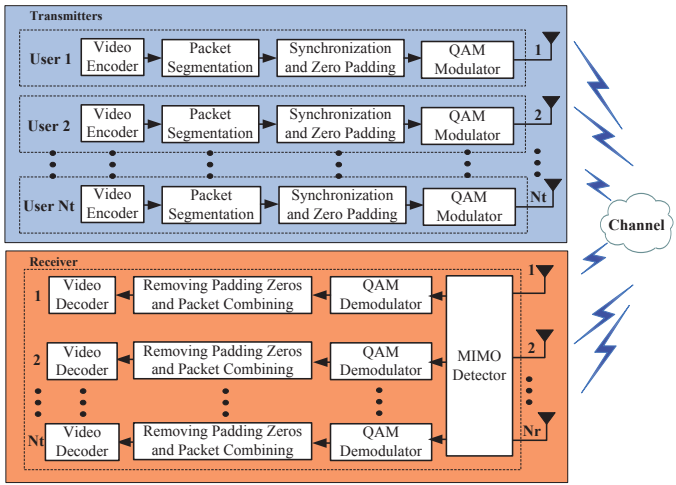}}
\caption{The proposed wireless video transmission architecture relying on LS-MIMO, where we have $N_r \gg N_t$. } \label{proposedsystem}
\end{figure}
Based on the above numerical evaluation of the BER performance as well as the discussions concerning their energy efficiency, in this section we propose a low-complexity M2P video transmission architecture relying on the uplink LS-MIMO system shown in Fig. \ref{proposedsystem}. 

At the transmit side, a standard video encoder, such as H.264\cite{Wiegand_2003:H_264_standard, Hanzo_2007:video_book} or H.265\cite{HEVC_2013:ITU, Sullivan_2012:HEVC_overview, Pourazad_2012:HEVC_magazine} may be first invoked at each user for video compression. Then, the output video streams of the video encoders are segmented into packets of different size for transport over IP networks.  In a common M2P wireless video transmission scenario, such as mobile video capture or multipoint wireless video monitoring, the users typically have a similar transmit power and rate. Therefore, with only a marginal loss of throughput, at the transmit side of the proposed architecture of Fig. \ref{proposedsystem} we use a zero-padding aided packet synchronisation mechanism, where a variable number of zeros are appended to the shorter packets, after completing the video encoding and packet segmentation. Explicitly, for packets of different size, the number of zeros appended by the zero-padding operation is different, so that after completing the zero-padding operation, all the packets would have the same size. This mechanism is capable of aligning video packets and synchronising the transmissions of multiple users. The bits in these aligned packets are then transmitted as QAM signals.  

At the receive side, firstly a MIMO detector is used for recovering the transmitted QAM symbols, which are then fed into QAM demodulators in order to obtain the bit streams of each user. The zeros appended to the shorter packets are removed and the resultant packets are combined for generating the original video streams. Finally, these video streams are fed into the video decoders of Fig. \ref{proposedsystem}, recovering the original video content of each user. 

For the small-scale MIMO benchmark  system, there will be a channel encoder immediately after the packet segmentation operation and correspondingly a channel decoder counterpart immediately before the video decoder.

\section{Performance Evaluation of HEVC/H.265 Video Transmission}
In this section, we benchmark the performance of the proposed LS-MIMO aided wireless video transmission system against that of the traditional small-scale MIMO assisted wireless video transmission system in Fig. \ref{psnrvssnr} and Fig. \ref{throughputvssnr}. 

\subsection{Peak Signal-to-Noise Ratio Performance Comparison}
\begin{figure*}[t]
\centering{\includegraphics[width=5in]{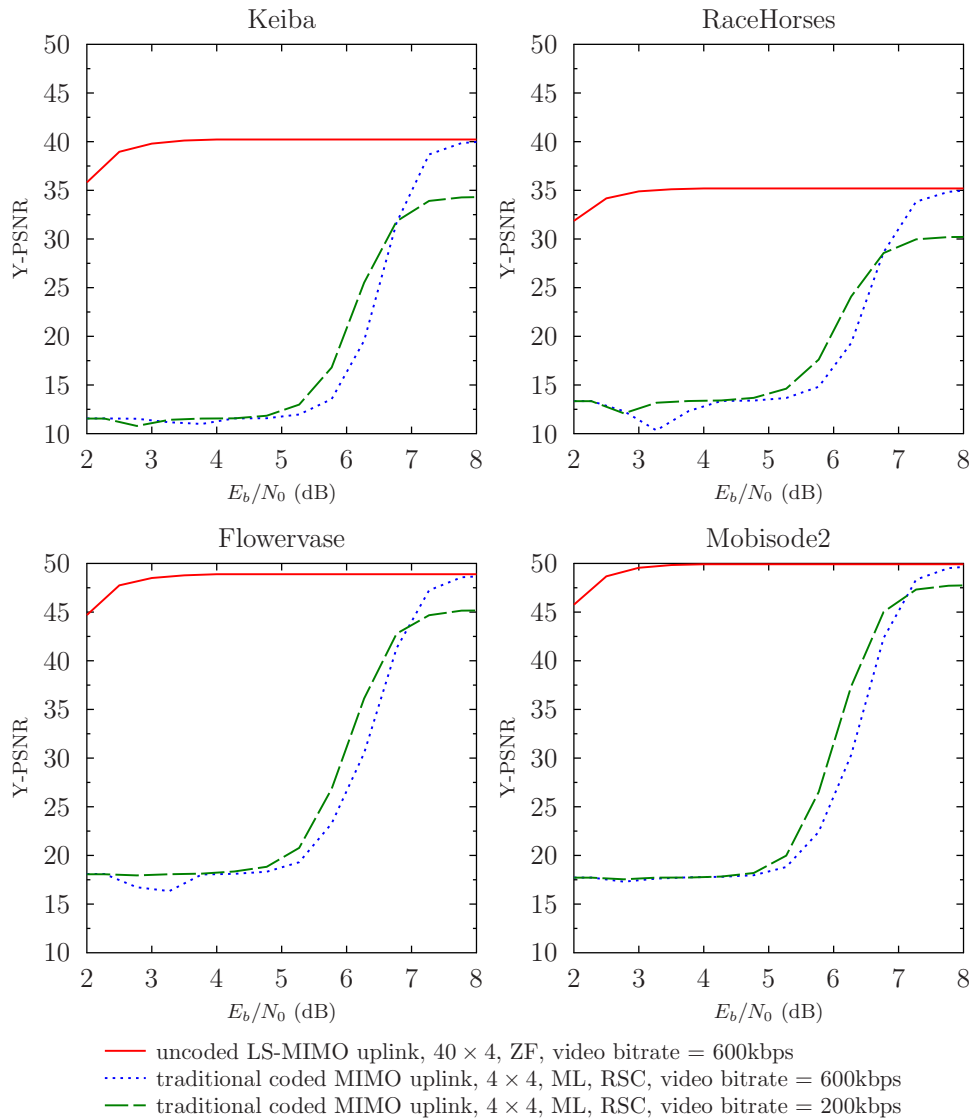}}
\caption{PSNR versus the channel-SNR $E_b/N_0$.} \label{psnrvssnr}
\end{figure*}
To benchmark the performance of the proposed LS-MIMO system against the traditional small-scale MIMO system, four Class D $[416\times240@30\text{Hz}]$ test sequences (Keiba, RaceHorses, Flowervase and Mobisode2) as required by the HEVC/H.265 evaluation\cite{Rosewarne_2014:software_HEVC} are invoked, each of which is transmitted by one of the four uplink users supported. All these standard sequences are provided by the Joint Collaborative Team on Video Coding (JCT-VC) expert group\cite{HEVC_2013:ITU, Rosewarne_2014:software_HEVC}. All sequences associated with different texture characteristics and motion activities are encoded using the HEVC/H.265 reference software (HM ver.13.1)\cite{Rosewarne_2014:software_HEVC}, assuming a frame rate of 30 frame/second (fps), $200$ frames and random access coding mode. 

Two scenarios are considered, since the transmission delay and source video quality constitute a pair of conflicting metrics to be traded off against each other for a given wireless transmission rate. First, when the same target video bitrate of $600 \text{kbps}$ is assumed for both schemes, implying that the quality of the input video streams for both schemes is the same, the error-free peak signal-to-noise ratio (PSNR) (Y component) of the above-mentioned four test sequences are $40.22\, \text{dB}$, $35.19\, \text{dB}$, $48.89\, \text{dB}$ and $49.92\, \text{dB}$, respectively. As a result, for the $(40\times 4)$-element LS-MIMO system and the $(4 \times 4)$-element conventional small-scale MIMO system that have the same \textit{raw transmission rate} in terms of the MIMO link, the transmission delay imposed by the conventional small-scale MIMO system may be three times as high as that of the proposed LS-MIMO system, because the $1/3$-rate RSC results in a threefold file-size increase. Therefore, this scenario is suitable for delay-tolerant video-file delivery applications, such as high-definition film downloading/uploading. Note that the ``three times transmission delay'' is estimated relying on the assumption that the processing delay per bit at the receivers of both systems is the same. However, as we mentioned in Section \ref{sec: case_study}, the receiver of the channel-coded small-scale MIMO benchmark system may have a higher computational complexity than our uncoded LS-MIMO scheme. Therefore, in practice, the transmission delay of the channel-coded small-scale MIMO benchmark system may be even higher than three times. On the other hand, 
for delay-sensitive interactive video applications, such as live video streaming, the delay specifications of the two systems are expected to be the same. In this case, the quality of the source video of the conventional $1/3$-rate RSC-coded small-scale MIMO system considered has to be degraded, hence the source video bitrate has to be reduced to $200 \text{kbps}$, which translates to the error-free PSNR (Y component) of $34.35\, \text{dB}$, $30.25\, \text{dB}$, $45.21\, \text{dB}$ and $47.80\, \text{dB}$ for the four test sequences considered, respectively.   

All video bit-streams obtained from the HEVC encoder are in form of the network-abstract-layer-units (NALUs), which are transmitted on a NALU basis as well. The NALU packets belonging to the same user maintain their natural chronological order. Each user has one NALU packet to be transmitted during each transmission interval and these packets are aligned in time to form a transmission block, which facilitates parallel transmission using the MIMO scheme advocated. Moreover, each NALU is protected by cyclic redundancy check (CRC) codes, whose official implementation is given in the reference software (HM ver.13.1) \cite{Rosewarne_2014:software_HEVC}. We note that in video communications, typically CRC codes are used for detecting whether a bitstream is error-free or not at the output of the channel decoder. This feature is supported by most video compression standards, such as H.264/Advanced Video Coding (AVC) \cite{Wiegand_2003:H_264_standard, Hanzo_2007:video_book} and H.265/HEVC\cite{HEVC_2013:ITU, Sullivan_2012:HEVC_overview, Pourazad_2012:HEVC_magazine, Rosewarne_2014:software_HEVC}. At the receiver, each channel-decoded/MIMO-detected NALU failing to pass the CRC process is removed during the packet combining process, namely prior to the video decoding. Then, each adjusted bitstream belonging to different users is sent to the individual video decoders, respectively, as shown in Fig. \ref{proposedsystem}.  

Additionally, a video frame-copy based two-stage error concealment mechanism is also invoked at the receiver. The reference codec of the previous video compression standard H.264/AVC provides an error concealment feature \cite{Wiegand_2003:H_264_standard}, which used frame-substitution for all the lost video pictures automatically. Unfortunately, this is no longer the case in the H.265/HEVC development \cite{HEVC_2013:ITU, Sullivan_2012:HEVC_overview, Pourazad_2012:HEVC_magazine, Rosewarne_2014:software_HEVC}. Our investigations carried out in this paper show that the H.265/HEVC codec only supports a basic level of error concealment. When a NALU packet containing a video picture is lost, the H.265/HEVC video decoder first attempts to conceal the picture loss by using its preceding reference picture as a direct frame-copy. But if the reference picture is not in the decoded-picture-buffer (DPB), the decoder becomes incapable of filling the lost picture. This indicates that not all the lost video pictures can be replenished by the H.265/HEVC video decoder. Therefore, a further complementary error concealment mechanism is implemented in the systems considered. More specifically, the lost pictures that cannot be replenished by the video decoder itself are filled in by using another frame-copy based error concealment mechanism, which copies the closest video picture in time instead of the reference picture to replace the lost picture. As a beneficial result, this two-stage error concealment mechanism is able to replenish all the lost pictures in the video decoding module.  In all of our experiments, each video bit-stream is transmitted 2000 times in order to generate statistically sound performance results.

More specifically, the PSNR versus $E_b/N_0$ (i.e. channel-SNR) performance of both the proposed LS-MIMO aided wireless video transmission system and the traditional small-scale MIMO assisted wireless video transmission system is presented in Fig. \ref{psnrvssnr}, where we observe that the proposed system substantially outperforms the traditional system, despite using no ECC for both the delay-tolerant and the real-time interactive video applications. Let us consider the Keiba sequence as an example. We can see that in terms of its video quality improvement the proposed scheme outperforms the traditional scheme by about $\Delta_\mathsf{PSNR} \approx 22.5\, \text{dB}$ at an  $E_b/N_0 = 6\, \text{dB}$ for a video bitrate of $600\text{kbps}$ (delay-tolerant video download applications), and about $\Delta_\mathsf{PSNR} \approx 20\, \text{dB}$ at an $E_b/N_0 = 6\, \text{dB}$ for a video bitrate of $200\text{kbps}$ (real-time lip-synchronized video applications). Alternatively, a channel-SNR gain of $\Delta_ {E_b/N_0}\approx 5\,\text{dB}$ is observed at a 
video PSNR of $36\, \text{dB}$ for the video bitrate of $600\text{kbps}$. Again, an even higher gain is achieved in the real-time lip-synchronized video application scenario, where the video bitrate of the conventional benchmarking scheme is $200\text{kbps}$.  

It should be noted that in the scenario of having a video bitrate of $600 \text{kbps}$, both systems have a similar PSNR versus channel-SNR performance in the high-SNR region. This is due to the fact that the NALU packet error rate (PER) is extremely low in the high-SNR scenario in both systems. However, for the scenario of real-time video applications, the PSNR achieved by the proposed LS-MIMO scheme is higher than that of the conventional small-scale coded MIMO scheme. This is because the video bitrate of the proposed LS-MIMO scheme is $600 \text{kbps}$, while the conventional small-scale MIMO scheme exhibiting the same transmission delay has a video bitrate of $200 \text{kbps}$.

\subsection{Throughput Performance Comparison} 
\begin{table}[t]
\begin{small}
\begin{center}
\renewcommand{\arraystretch}{1.3}
\caption{Parameters of an LTE-like system using 4-QAM} \label{ffdltepar}
{\begin{tabular}{|l|l|}\hline
\textbf{Parameters} & \textbf{Value} \\\hline
Bits per modulation symbol & $M_{\text b} = \log_2M = 2$ \\\hline
Number of transmit antennas  & $N_{\text{t}} = 4$ \\\hline
Number of receive antennas & $N_{\text r} =\{4, 40\}$ \\ \hline
Duration per time slot & $T_{\text{ts}} = 0.5 \text{ms}$ \\\hline
Number of orthogonal frequency-division &  $M_{\text{spts}} = 7$ \\
multiplexing (OFDM) symbols per time slot &  \\\hline
Total bandwidth & $B = 20 $MHz \\ \hline
Total number of resource blocks (RBs) for  & $M_{\text{RB}} = 100$  \\ 
data transmission per time slot & \\ \hline
Number of subcarriers per RB & $M_{\text{f}} = 12$  \\ \hline
Channel coding rate & $R_{\text c} = \{1, 1/3\}$ \\ \hline
\end{tabular}}{} 
\end{center}
\end{small}
\end{table}

\begin{figure}[t]
\centering{\includegraphics[width=\linewidth]{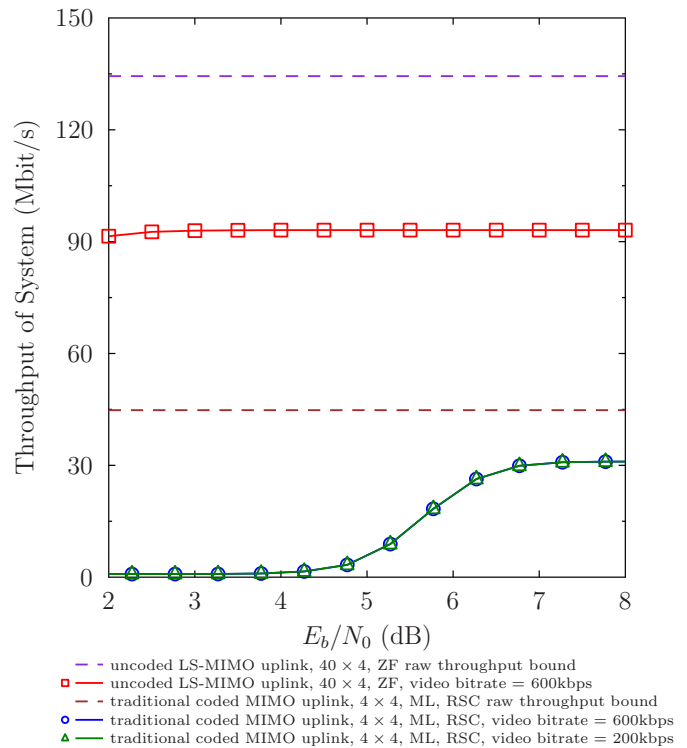}}
\caption{Throughput versus the channel-SNR $E_b/N_0$.} \label{throughputvssnr}
\end{figure}
For further demonstration of the performance discrepancy between the two systems, let us calculate the attainable throughput using the parameters of a 4QAM-based Long Term Evolution (LTE)-like system, as shown in Table \ref{ffdltepar} (see J. Zyren and W. McCoy, ``Overview of the 3GPP long term evolution physical layer'', Freescale Semiconductor, Inc., Jul., 2007 for more details). The \textit{theoretical raw throughput} of the systems assuming perfectly error-free channel decoding may be formulated based on the resource block (RB) and frame structure defined in LTE as follows:  
\begin{equation}\label{eq:theoretical_throughput_calculation}
C_{\text{raw}} = \frac {M_{\text b} \times M_{\text{spts}} \times M_{\text{RB}} \times M_{\text{f}} \times N_{\text t} \times R_{\text c}} {T_{\text{ts}}}, 
\end{equation}
where the channel coding rate is set to $R_{c} = 1 $ in the proposed system, since no channel codec is used. To be more specific, (1) is essentially the number of information bits transmitted in a time slot divided by the duration of the time slot. Relying on Eq. \eqref{eq:theoretical_throughput_calculation}, the theoretical raw throughput bound of the proposed uncoded LS-MIMO system is $C_{\text{raw, LS}} = 134.4\text{Mbps}$. By contrast, the theoretical bound of the raw throughput of the conventional benchmark MIMO system using a rate-$1/3$ channel codec is $C_{\text{raw,B}} = 44.8 \text{Mbps}$. In other words, the theoretical raw throughput of the propsoed scheme is three times as high as that of the benchmark scheme.

For the sake of visualizing the achievable effective throughput of the two systems, we define $\mathsf{BER}_{\text{PPR}}(E_b/N_0)$ as the post-packet-recovery (PPR) error rate, which is calculated as
\begin{equation} \label{BER_ppr}
\mathsf{BER}_{\text{PPR}}(E_b/N_0) = \frac{N_{b,\text{PPR}}}{N_{b,t}},
\end{equation}
where $N_{b, \text{PPR}}$ is the total number of erroneous bits after PPR, while $N_{b,t}$ is the total number of bits transmitted by the four users supported.  Therefore, the achievable effective throughput of both the systems is given by      
\begin{equation}\label{eq:effective_throughput_calculation}
C_{\text{eff}}(E_b/N_0) = C_{\text{raw}} \times \left[1 - \mathsf{BER}_{\text{PPR}}(E_b/N_0) \right].
\end{equation}

Finally, the achievable effective throughput of the proposed architecture and of the conventional benchmark architecture is shown in Fig.~\ref{throughputvssnr}. Observe in Fig. \ref{throughputvssnr} that neither of the systems are capable of reaching the theoretical bound. The reasons for this observation are that 1) both systems suffer from the channel-induced transmission errors and 2) about $30$ bits are wasted in the zero padding of each NALU in the propose scheme. However, the effective throughput of the proposed scheme is still at least twice higher than that of the benchmark system at most channel-SNR values of interest, which is primarily because the latter invokes a rate-$1/3$ channel codec. Additionally, it is worth noting that the achievable effective throughput of the benchmark system supporting the video bitrates of $600 \text{kbps}$ and of $200 \text{kbps}$ remains  identical, which is because the transmission capability of a wireless communication 
system is independent of the information source. 

\section{Conclusions}
In this paper, a transmission architecture relying on the emerging large-scale MIMO (LS-MIMO) technique was proposed for future wireless video communications. We demonstrate that for the most advanced HEVC/H.265 video codec, the proposed architecture invoking the low-complexity linear ZF detector and no channel coding may substantially outperform the conventional small-scale MIMO based architecture employing the high-complexity optimal ML detector and a rate-$1/3$ RSC codec in terms of the achievable effective throughput as well as in terms of the reconstructed video quality (PSNR) for both delay-tolerant and real-time video applications.

\ifCLASSOPTIONcaptionsoff
  \newpage
\fi

\bibliography{../../../shaoshi_bib/IEEEfull,../../../shaoshi_bib/shaoshi_reference}

\begin{IEEEbiography}[{\includegraphics[width=1in,height=1.25in,clip,keepaspectratio]{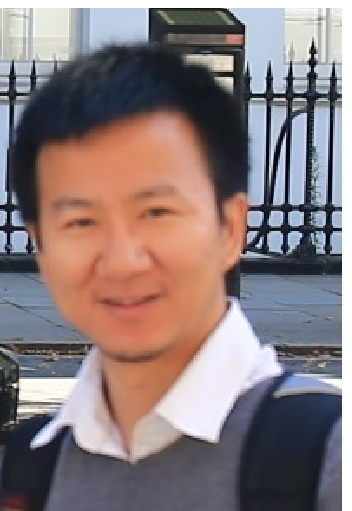}}] {Shaoshi Yang}
(S'09-M'13) received his B.Eng. degree in Information Engineering from Beijing University of Posts and Telecommunications (BUPT), Beijing, China in Jul. 2006, his first Ph.D. degree in Electronics and Electrical Engineering from University of Southampton, U.K. in Dec. 2013, and his second Ph.D. degree in Signal and Information Processing from BUPT in Mar. 2014. He is now working as a Postdoctoral Research Fellow in University of Southampton, U.K. From November 2008 to February 2009, he was an Intern Research Fellow with the Communications Technology Lab (CTL), Intel Labs, Beijing, China, where he focused on Channel Quality Indicator Channel (CQICH) design for mobile WiMAX (802.16m) standard. His research interests include MIMO signal processing, green radio, heterogeneous networks, cross-layer interference management, convex optimization and its applications. He has published in excess of 25 research papers on IEEE journals. 

Shaoshi has received a number of academic and research awards, including the prestigious Dean's Award for Early Career Research Excellence at University of Southampton, the PMC-Sierra Telecommunications Technology Paper Award at BUPT, the Electronics and Computer Science (ECS) Scholarship of University of Southampton, and the Best PhD Thesis Award of BUPT. He is a member of IEEE/IET, and a junior member of Isaac Newton Institute for Mathematical Sciences, Cambridge University, U.K. He also serves as a TPC member of several major IEEE conferences, including \textit{IEEE ICC, GLOBECOM, VTC, WCNC, PIMRC, ICCVE, HPCC}, and as a Guest Associate Editor of \textit{IEEE Journal on Selected Areas in Communications.} (\url{https://sites.google.com/site/shaoshiyang/}) 
\end{IEEEbiography}

\begin{IEEEbiography}[{\includegraphics[width=1in,height=1.25in,clip,keepaspectratio]{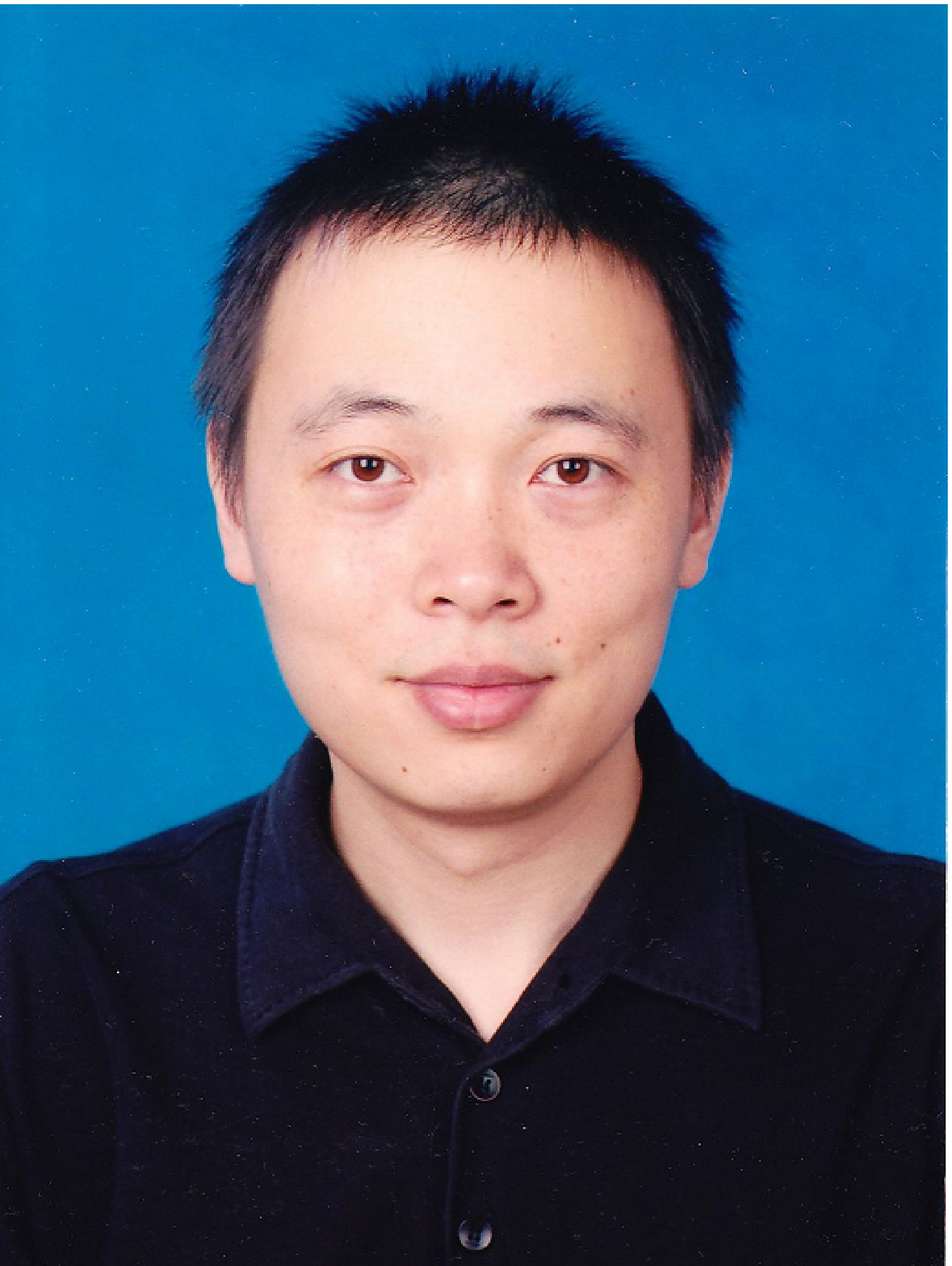}}] {Cheng Zhou} received the B.S. degree in Electronics and Information Engineering from Hubei University, Wuhan, China, in 2001, and the M.S. degree in Pattern Recognition and Intelligent Systems, and the Ph.D. degree in Control Science and Engineering from Huazhong University of Science and Technology, Wuhan, China, in 2007 and 2010, respectively. He is currently a Lecturer with the School of Electronics and Information Engineering, South-Central University for Nationalities (SCUEC), Wuhan, China. He was a Visiting Scholar with the School of Electronics and Computer Science, University of Southampton from Nov. 2013 to Oct. 2014. He has been involved in the research of multimedia signal processing, particularly, video coding for MPEG-4, H.264, AVS (China) and HEVC standards since 2004.
\end{IEEEbiography}

\begin{IEEEbiography}[{\includegraphics[width=1in,height=1.25in,clip,keepaspectratio]{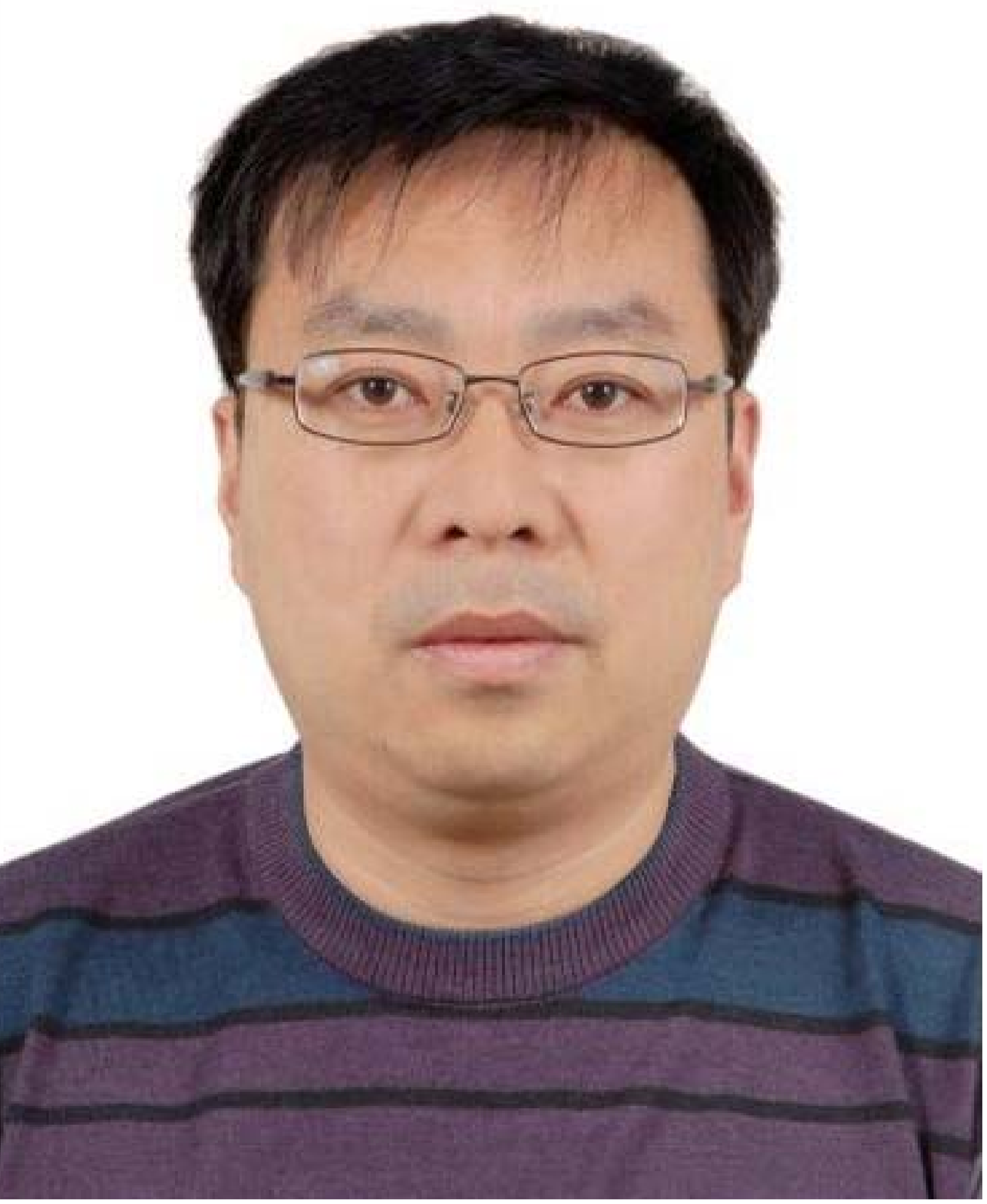}}] {Tiejun Lv}
(M'08-SM'12) received the M.S. and
Ph.D. degrees in electronic engineering from the
University of Electronic Science and Technology
of China (UESTC), Chengdu, China, in 1997 and
2000, respectively. From January 2001 to January
2003, he was a Postdoctoral Fellow with Tsinghua
University, Beijing, China. In 2005, he became a Full
Professor with the School of Information and Communication
Engineering, Beijing University of Posts and Telecommunications
(BUPT). From September 2008 to March 2009, he was a Visiting Professor with
the Department of Electrical Engineering, Stanford University, Stanford, CA. 
He is the author of more than 200 published technical papers on
the physical layer of wireless mobile communications. His current research
interests include signal processing, communications theory and networking.
Dr. Lv is also a Senior Member of the Chinese Electronics Association.
He was the recipient of the Program for New Century Excellent Talents in
University Award from the Ministry of Education, China, in 2006.
\end{IEEEbiography}

\begin{IEEEbiography}[{\includegraphics[width=1in,height=1.25in,clip,keepaspectratio]{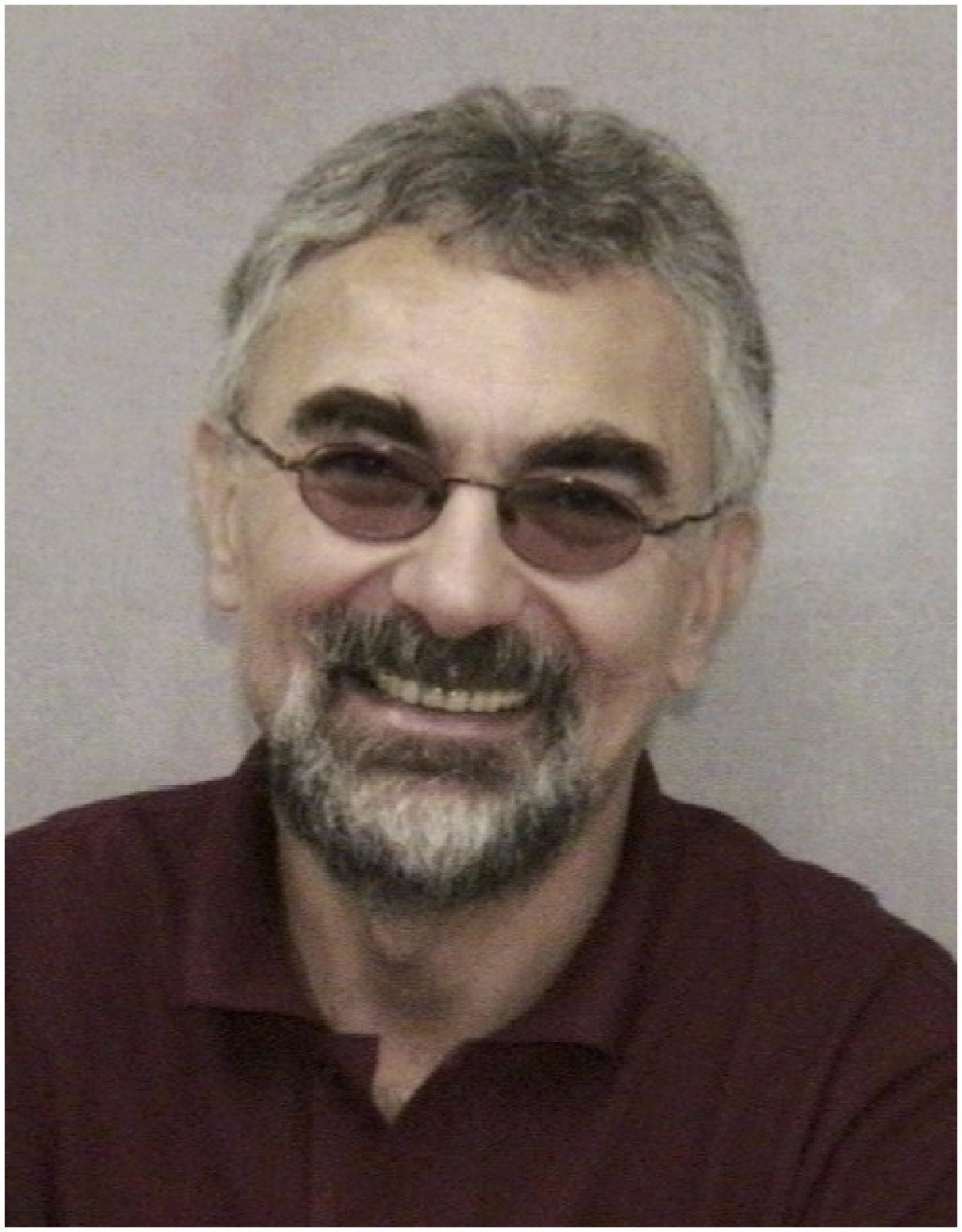}}] {Lajos Hanzo}
(M'91-SM'92-F'04) received his degree in electronics in
1976 and his doctorate in 1983. He was awarded an honorary
doctorate by the Technical University of
Budapest in 2009 and by the University of Edinburgh in 2015.  During his 39-year career in telecommunications he has held
various research and academic posts in Hungary, Germany and the
UK. Since 1986 he has been with the School of Electronics and Computer
Science, University of Southampton, UK, where he holds the Chair in
Telecommunications.  He has successfully supervised about 100 PhD students,
co-authored 20 John Wiley/IEEE Press books on mobile radio
communications totalling in excess of 10 000 pages, published 1500+
research entries at IEEE Xplore, acted both as TPC and General Chair
of IEEE conferences, presented keynote lectures and has been awarded a
number of distinctions. Currently he is directing a 60-strong
academic research team, working on a range of research projects in the
field of wireless multimedia communications sponsored by industry, the
Engineering and Physical Sciences Research Council (EPSRC) UK, the
European Research Council's Advanced Fellow Grant and the Royal
Society's Wolfson Research Merit Award.  He is an enthusiastic
supporter of industrial and academic liaison and he offers a range of
industrial courses. 

He is also a Fellow of the Royal Academy of Engineering, of the Institution
of Engineering and Technology (IET), and of the European Association for Signal
Processing (EURASIP). He is a Governor of the IEEE VTS.  During
2008 - 2012 he was the Editor-in-Chief of the IEEE Press and a Chaired
Professor also at Tsinghua University, Beijing. He 
has 24 000+ citations. For further information on research in progress and associated
publications please refer to \url{http://www-mobile.ecs.soton.ac.uk} 
\end{IEEEbiography}

\end{document}